\documentclass[reprint, amsmath,amssymb, aps, superscriptaddress,prl]{revtex4-2}

\usepackage{graphicx}
\usepackage{dcolumn}
\usepackage{bm}
\usepackage{amsthm}
\usepackage{hyperref}
\usepackage[mathlines]{lineno}
\usepackage{geometry}
\begin{document}

\title{Efficient Transport Networks through Edge Pruning by Relative Resistance}

\author{Mathieu Le Verge-Serandour}
\affiliation{%
 School of Natural Science, Technical University of Munich
}%

\author{Viktor Vasilev}
\affiliation{%
 School of Natural Science, Technical University of Munich
}%

\author{Lara Szeimies}
\affiliation{%
 School of Natural Science, Technical University of Munich
}%

\author{Karen Alim}
\affiliation{%
 School of Natural Science, Technical University of Munich
}%
\email{k.alim@tum.de}

\begin{abstract} 
Transport networks are key to living and engineered systems. How can transport networks adapt efficiently using only local rules, without a global objective or specific boundary conditions? Inspired by observations in biological transport networks, we adapt networks by subsequently pruning edges by relative resistance, thus removing edges whose resistance is higher than the resistance for routing flow through the rest of the network. We find that adaptation by such a local, boundary-condition-free measure of redundancy, set by topology alone, yields sparse, loop-rich, low-dissipation networks.
\end{abstract}

\date{\today}

\maketitle

Transport networks are key to moving resources over distances in both living and engineered systems, where mere diffusion would be too slow. Yet, each edge in a transport network is costly; therefore, the question of how to devise efficient transport networks has attracted computer scientists \cite{Spielman2011a, Forrow2018}, applied mathematicians \cite{Kaiser2020}, and physicists \cite{Bohn2007, Corson2010, Katifori2010, Cai2023a, Kirkegaard2020} alike.

Biological transport networks, shaped by the pressure of evolution, provide inspiration for solving the problem. Our blood vasculature, our pancreatic ducts \cite{DahlJensen2018}, plants \cite{Katifori2018}, fungi \cite{Fricker2017a}, or the slime mold \textit{Physarum polycephalum} \cite{Tero2010,Alim2018} adapt their network architecture by adjusting edge resistances, eventually pruning individual edges. The seminal work of Murray a century ago formalized vasculature adaptation by minimizing flow dissipation at finite building cost for network edges \cite{Murray1926}. Minimization of this global cost function predicts optimal network architectures \cite{Bohn2007, Durand2007, Corson2010, Katifori2010}, however, global knowledge of the cost function is not required. Instead, optimal networks also follow from a local rule based on the fluid flow, where edge resistance shrinks at large flow, when the edge is crucial for transport, and grows at low flow even to the extend of edge pruning, when the edge is superfluous \cite{Hu2013, Ronellenfitsch2016, Ronellenfitsch.20190t}. Independent of the optimization, the final architecture depends on the imposed boundary conditions (external pressures or flows, fixed or fluctuating) since the locations of sources and sinks strongly bias which network edges ultimately persist. 
Additional factors, such as molecular transport \cite{Meigel2018a, Kramer2023}, erosion dynamics \cite{Zareei2022, Bouvard2025}, source/sink correlations \cite{Ronellenfitsch.20190t}, or transport time \cite{Kirkegaard2020}, further enrich network behavior, giving rise to memory imprinting \cite{Bhattacharyya.2022}, learning \cite{Stern2025, Anisetti2023}, or resilience \cite{Gavrilchenko2019}.
Yet, existing models mostly describe adaptation as an optimization process, either global or local, prescribed by boundary conditions that strongly constrain the final architecture. 
This leaves a fundamental question open: how can a network reach and maintain an efficient, functional architecture from local rules alone, without a global objective or constraint by specific boundary conditions?

\begin{figure}[h!]
    \centering
    \includegraphics[width=\linewidth]{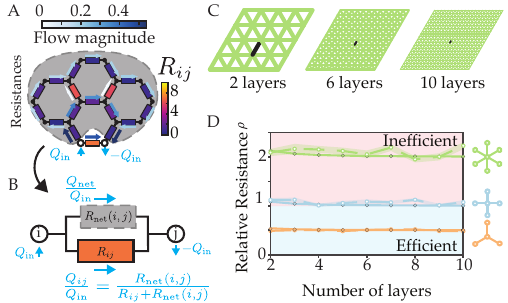}
    \caption{\textbf{Among regular networks, only hexagonal support low relative resistance everywhere}. (A) Relative resistance $\rho_{ij}$ measures the flow fraction through an edge $(i,j)$ versus the rest of the network if flow is injected at source $i$ and again withdrawn at sink $j$, and (B) follows from the ratio of edge resistance $R_{i,j}$ to the resistance of the rest of the network in an equivalent circuit $R_{\rm{net}}(i, j)$. (C) Calculating the relative resistance for the central edge (black) in, here, triangular networks, with increasing network size shows that (D) a central edge's relative resistance converges to an analytically predicted value, where only hexagonal networks are always efficient ($\rho<1$) while square are marginal ($\rho=1$), and triangular are inefficient ($\rho>1$). Solid lines indicate the analytical prediction, full circles correspond to regular networks, dashed and white circles to noisy resistances ($R_{ij}=R_0\pm\varepsilon$, $R_0=1$, $\varepsilon=0.3$).}
    \label{fig1}
\end{figure}

Again, biological transport networks inspire. The vein networks of \textit{P.~polycephalum} adapt as well following a local rule based on the flow \cite{Marbach.2023}. Yet, network adaptation happens in the absence of specific external boundary conditions, since flows are generated in a distributed manner by the local, rhythmic pumping of each network's edges \cite{LeVergeSerandour2024}. Therefore, the flow $Q_{ij}$ along an edge $(i,j)$ is a result of the flow $Q_{\rm{in}}$, generated by its fellow edges, injected at say edge endpoint $i$, acting as a source, and extracted at the other endpoint $j$, acting as a sink \cite{Marbach.2023}. But a fraction of the injected flow could also be routed through the rest of the network instead of through edge $(i,j)$. The partitioning of flow is set by the relative resistance of edge $(i,j)$, denoted $\rho_{ij}$, which compares the resistance $R_{ij}$ of said edge, to the equivalent resistance of the rest of the network without the edge, $R_{\rm{net}}(i, j)$, as $\rho_{ij}=R_{ij} / R_{\rm{net}}(i,j)$, see Fig.~\ref{fig1}A, B, specifying the flow along edge $(i,j)$ to $Q_{ij}/Q_{\rm{in}} \equiv 1/(1+\rho_{ij})$.  Thus, for large relative resistance $\rho_{ij}>1$, the edge's flow is low compared to the flow along alternative routes through the network and the edge is superfluous \cite{Marbach.2023rjf}. Crucially, although $\rho_{ij}$ is shaped by the network's entire topology through $R_{\mathrm{net}}(i,j)$, it is read out locally: an edge only needs to sense the flow it itself carries, without any knowledge of the network's global architecture.

Here, we investigate transport network adaptation by subsequently pruning edges of large relative resistance, $\rho_{ij}>1$, until no superfluous edges persist, $\rho_{ij}\leq 1$. We show that regular degree-three networks support low relative resistance everywhere, a state that can be obtained for any regular, random or non-planar network through subsequent edge pruning. We demonstrate that sequential pruning based only on local relative resistance drives networks toward efficient, sparse, loop-rich architectures without prescribing boundary conditions or minimizing a global objective.

We start by investigating how network topology, represented by node degree, affects the relative resistance $\rho$. As simple models of embedded flow, or equivalently, resistor networks, we consider three canonical planar regular networks with identical edge resistances: hexagonal ($\nu=3$), square ($\nu=4$), and triangular ($\nu=6$) lattices, where $\nu$ denotes the node degree. 
For each topology, we compute the relative resistance $\rho$ of a central edge as the network size increases (Fig.~1C), using Kron reduction \cite{Dorfler2013} to obtain $R_{\rm{net}}$ (see Supplementary Information, \cite{suppinfo}). In all cases, $\rho$ converges toward an individual, topology-dependent limiting value, which is only smaller than one for hexagonal networks and marginal for square networks. To reconcile the differing relative resistances, we express Foster's theorem for planar regular networks \cite{Ghosh2008} in terms of relative resistances, $\sum_{ij}\frac{1}{1+\rho_{i,j}}=n-1$, where $n$ is the number of nodes. As $\rho$ is uniform in an infinite, homogeneous network, we simplify the sum to a product $m\frac{1}{1+\rho_{ij}}$ with the number of edges, $m$. Finally, since for regular lattices, the number of edges and node degree are related by $m=\nu n/2$, we derive $\rho = \frac{\nu n}{2(n-1)}-1$ (see Supplementary Information, \cite{suppinfo}). In the large network limit ($n\gg1$), this expression simplifies to $\rho \sim (\nu-2)/2$ in agreement with our numerical results for the asymptotic relative resistance (Fig.~\ref{fig1}D).
Our analytical prediction is robust under the addition of weak multiplicative noise to the edges' resistances $R_{ij}=R_0(1+\epsilon)$, $\mathrm{Var}(\epsilon)=\sigma^2\ll1$, and holds true for periodic networks (see Supplementary Figure 1, \cite{suppinfo}). We find that node degree robustly determines relative resistance, and only hexagonal networks are efficient in terms of low relative resistance $\rho<1$, whereas square networks are marginal, $\rho\approx1$ but triangular networks are inefficient, $\rho>1$. The dependence on node degree suggests that pruning individual edges in high node degree networks might render them more efficient.

\begin{figure}[h!]
    \centering
    \includegraphics[width=\linewidth]{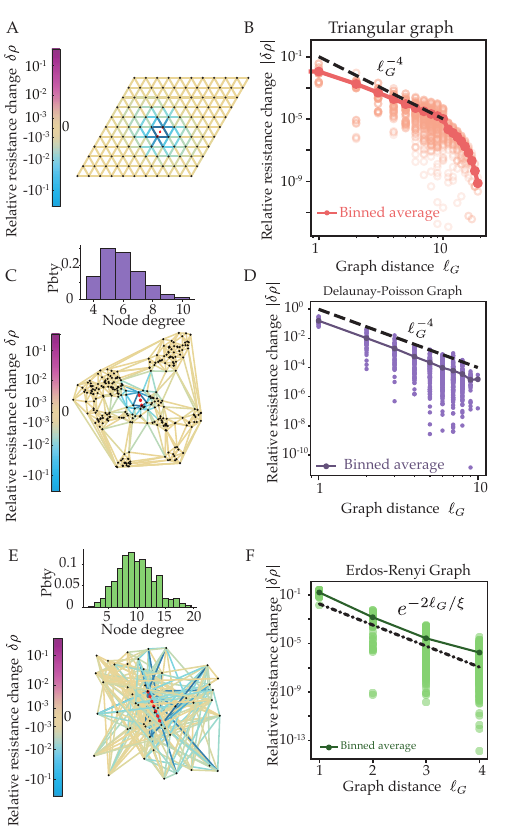}
    \caption{\textbf{Edge pruning locally decreases relative resistance in a quadrupole field-like manner}. Pruning an edge (red dashed line) imprints a quadrupole pattern in the change in relative resistance $\delta \rho$ for a (A) triangular network (5 layers) and a (C) Delaunay-Poisson network, but exhibits shallower decay for a (E) Erdős–Rényi network ($N=50, p=0.2$). The absolute relative resistance change for each edge versus network distance $\ell_G$ to the pruned edge decays in a power-law for (B) regular (10 layers) and random (C) planar-network, but exponentially for (F) Erdős–Rényi network, as predicted analytically (power-law as guide to the eye). (C, E) Inset: initial node degree distribution.}
    \label{fig2}
\end{figure}

We, therefore, next address how pruning edges, thereby reducing the node degree of the network, can reduce the relative resistance and improve the efficiency of a network.
Mathematically, pruning an edge $(i,j)$ corresponds to a rank-one update of the network Laplacian $L$, such that the updated Laplacian is $L'=L-\frac{1}{R_{ij}}\vec{b}_{ij}\otimes \vec{b}_{ij}^{T}$, with $\vec{b}_{ij}=\vec{e}_i-\vec{e}_j$ selecting the pruned edge by the basis vectors of its nodes $\vec{e}_{i},\, \vec{e}_{j}$. We derive that this update changes the relative resistance, $\delta \rho_{uv} \equiv \rho'_{uv} - \rho_{uv}$, of all remaining edges $(u,v)$ through the Green's function $G=L^\dagger$, the pseudo-inverse of the Laplacian $L$ (see Supplementary Information, \cite{suppinfo}), by
\begin{equation}
\label{eq:change_rho_removal}
\delta \rho_{uv} = -\frac{R_{uv}}{R_{\rm{eff}}(u,v)}\frac{\gamma_{ij}c^2_{uv,ij}}{R_{\rm{eff}}(u,v)+\gamma_{ij}c^2_{uv,ij}},
\end{equation}
where $\gamma_{ij}= \frac{1}{\rho_{ij}R_{\rm{eff}}(i,j)}$ is defined via the effective resistance $R_{\rm{eff}}(u,v)\equiv \vec{b}^T_{uv}L^\dagger\vec{b}_{uv}$ between nodes $u$ and $v$, a well-known quantity in graph theory \cite{Spielman2011a, Ghosh2008, Tetali1991,Klein1993}, related to the relative resistance as $\rho_{ij} = R_{ij}/R_{\rm{eff}}(i,j) - 1$ (see Supplementary Information, \cite{suppinfo}). 
We, further, define the edge--edge coupling function $c_{uv,ij}=\vec{b}^T_{uv} L^\dagger \vec{b}^T_{ij}$, which can be understood as a discrete second derivative of the lattice Green's function. From the analytical result, we directly read off that edge pruning always reduces the relative resistance in the remaining edges, though the impact decays, due to the nature of the lattice Green's function, for planar networks in a coarse-grained, discrete quadrupole-like pattern (Fig.~\ref{fig2}A), similar to a dipole current source created by network damage \cite{Gavrilchenko2019}. The decay is driven by the edge--edge coupling function $c_{uv,ij}$, which we evaluate in the continuum limit, where for planar networks the Green's function can be expressed as $G(\ell_G)\sim \log(\ell_G)$ \cite{Manik2017}, in terms of the graph distance $\ell_G$ between edges $(u,v)$ and $(i,j)$, which corresponds to the shortest-path distance between the closest pair of nodes belonging to edges $(u,v)$ and $(i,j)$. Translated to the change in relative resistance, we predict the impact of pruning an edge to decay as $\delta\rho(\ell_G)\sim -\ell_G^{-4}$ for planar networks (Fig.~\ref{fig2}B) in agreement with numerical results, see Supplementary Information, \cite{suppinfo}. Periodic regular networks also confirm the scaling (Supplementary Figure 2, \cite{suppinfo}). 

Inspired by our analytical insight into the relative resistance decay due to pruning, we test our scaling on a random planar network with a comparable node degree, generated by Delaunay triangulation of Poisson-distributed points (Fig.~\ref{fig2}C). We confirm the decay of relative resistance (Fig.~\ref{fig2}D), despite the variations in node degree and spatial node clustering compared to the regular networks. Some edges, however, located with low connectivity to the network part of the pruned edge, seem to be less affected (Fig.~\ref{fig2}C).
In random Erdős–Rényi networks (Fig.~\ref{fig2}E), we find instead a shallower decay that we derive to be exponential with the network distance $\ell_G$, $\delta \rho(\ell_G) \sim -\exp(-2\ell_G/\xi)$ (Fig.~\ref{fig2}F) in agreement with numerical solutions; (see Supplementary Information, \cite{suppinfo})
Therefore, edge pruning leads to a local decrease in relative resistance, consistent with the fact that changes in topology primarily affect the immediate vicinity of the network. Thus, we deduce that sequential pruning is necessary to make networks efficient, i.e., to remove all superfluous edges such that all remaining edges have $\rho_{ij}\leq 1$.

\begin{figure}
    \centering
    \includegraphics[width=1\linewidth]{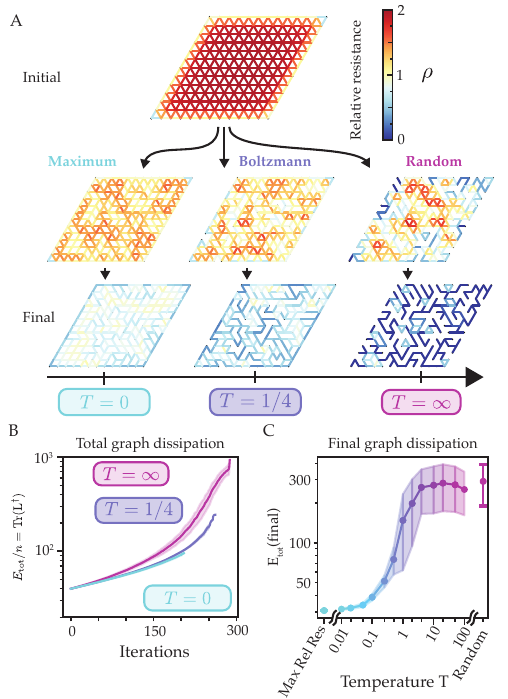}
    \caption{\textbf{Sequential pruning of maximum relative resistance edges leads to low dissipation networks}. Starting from a triangular lattice (6 layers), we prune edges (A) selected by maximum relative resistance ($T=0$), with Boltzmann probabilities ($T=\frac{1}{4}$) or at random ($T=\infty$) until all edges $\rho_{ij}\leq 1$. Patterns of relative resistance during adaptation and in final networks differ. (B) The total dissipation of the networks during pruning adaptation across schemes reveals that maximum pruning evolves at lower dissipation with fewer pruning. Standard deviation is shown in semi-opaque regions. (C) The final dissipation $E_{\rm{tot}}(\rm{final})$ for different temperatures $T$ shows that schemes behave like maximum or random pruning when the temperature is far from $T=1$. For each case, 50 simulations of 5 layer networks were averaged.}
    \label{fig3}
\end{figure}

We therefore introduce a network adaptation scheme, where superfluous edges with $\rho_{ij}>1$ 
 are sequentially removed until all remaining edges satisfy $\rho_{ij}\leq1$. Starting from a regular triangular loopy network,  
we compare three schemes: (i) maximum pruning, where the edge of maximum relative resistance is removed, (ii) random pruning, where an edge is chosen randomly provided it does not disconnect the network, and (iii) Boltzmann-weighted pruning, an intermediate between random and maximum pruning. Here, an edge $(i,j)$ is removed with probability $p_{ij}=\frac{\exp\left(\frac{\rho_{ij}-\rho_{\max}}{T}\right)}{\sum_{\langle kl \rangle \in E} \exp \left(\frac{\rho_{kl}-\rho_{\max}}{T}\right)}$, where the temperature-like control parameter $T$, recovers the maximum scheme for $T=0$, and the random one for $T=\infty$. We observe that more and more edges need to be pruned to achieve $\rho_{ij}\leq1$ for all edges as $T$ increases (Fig.~\ref{fig3}A). Pruning more edges reduces the building cost of the overall network more. But then pruning edges increases the effective resistance and thereby dissipation by Rayleigh's monotonicity law \cite{Klein1993}. In addition, the final relative resistance patterns vary for the adaptation dynamics of the relative resistance distribution, see Supplementary Figure 3, \cite{suppinfo}. It is therefore not clear which adaptation scheme yields efficient networks.

To evaluate a networks' efficiency for any choice of boundary condition, we sum over the dissipation within each single edge $(i,j)$ and all permutations of source node $s$ and sink node $t$, $E_{\rm{tot}} =\sum_{s<t}\sum_{ij} R_{ij}q_{ij}^2(s,t)$, which simplifies to $E_{\rm{tot}}=n\,\operatorname{Tr}(L^\dagger)$\cite{Ghosh2008} (see Supplementary Information, \cite{suppinfo}).
We observe that the total dissipation is always lowest for maximum pruning (Fig.~\ref{fig3}B, Supplementary Figure 4). In fact, total dissipation increases with Boltzmann temperatures in a sigmoid/logistic function, which saturates at $T=0$ and $T=\infty$, with an inflection point located around $T=1$, showing that for relative resistance pruning $T<1$, the total dissipation is closer to the maximum scheme, while total dissipation for $T\geq1$ appears similar to the random scheme (Fig.~\ref{fig3}C). Thus, sequential edge pruning by relative resistance consistently outcompetes the alternative adaptation schemes to yield low dissipation, efficient networks. 

\begin{figure}[h!]
    \centering
    \includegraphics[width=1\linewidth]{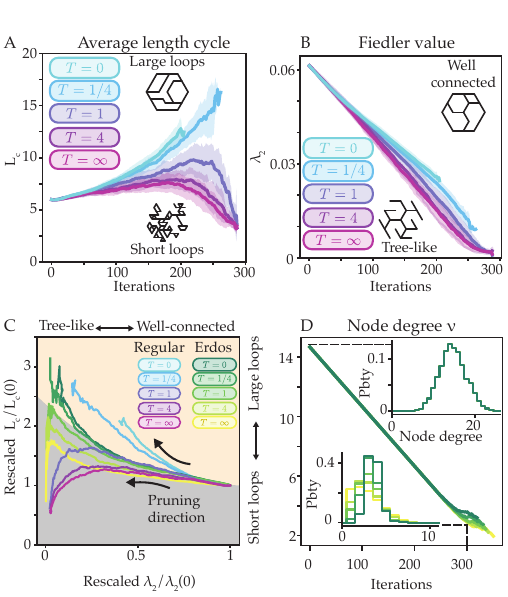}
    \caption{\textbf{Pruning maximum relative resistance edges homogenizes network topology and preserves loop structures.}  (A) Sequentially pruning edges shows a split in the average length cycle $L_c$, with relative resistance pruning ($T<1$) converging towards large loopy structures while rather random pruning ($T\geq1$) converges towards short loops. (B) Quantifying network connectivity with the Fiedler value $\lambda_2$ shows that random pruning converges faster to tree-like structures with low $\lambda_2$. (C) Combining both measures into a morphology diagram shows two regions emerging: random pruning $T\geq1$ converges to tree-like networks with short loops, while relative resistance pruning $T<1$ leads to networks with larger loops that are more connected. This applies to Erdős–Rényi random networks as well, where (D) the maximal relative resistance pruning homogenizes the node degree $\nu$. Top right inset: initial distribution. Bottom left inset: final distribution. We used 6 layers for the regular networks, and $N=50, p=0.3$ for the Erdős–Rényi networks, resulting in roughly 400 edges each; $\langle R_{ij}(0)\rangle=1$, $\langle R^2_{ij}(0)\rangle=0.3$. Ensemble average over 50 simulations.}
    \label{fig4}
\end{figure}

The resulting pruned networks exhibit visually distinct topological configurations, see (Fig.~\ref{fig3}A), with relative resistance pruning preserving many loops and connectivity, and random pruning generating separated tree-like structures. Indeed, the average cycle length 
 $L_c$ monotonously increases for relative resistance pruning ($T<1$) toward large cycles. On the contrary, in random pruning ($T\geq 1$) average cycle length eventually decreases toward tree-like structures with residual short cycles (Fig.~\ref{fig4}A). Quantifying network connectivity by the second smallest eigenvalue of the Laplacian, $\lambda_2$, the so-called Fiedler value \cite{Ghosh2008, Manik2017,Fiedler1973}, underlines the loss in connectivity with the Fiedler value reaching near zero values (Fig.~\ref{fig4}B) for random pruning, but higher values for maximum relative resistance pruning. We rationalize the difference in $\lambda_2$ decay by deriving that the change in Fiedler value due to pruning an edge $(i,j)$ depends inversely on the relative resistance of the pruned edge $\rho_{ij}$: $\delta \lambda_2\sim -\frac{\lambda_2}{1+\rho_{ij}}+O(\frac{\lambda_2}{ R_{ij} \lambda_3})$ (see Supplementary Information, \cite{suppinfo}). So pruning of maximum resistance edges $\rho_{\max} > \rho_{ij}$ results in a shallower change in connectivity than in any other pruning scheme. Our observation on larger average cycle length and higher connectivity with relative resistance pruning generalizes to random Erdős–Rényi networks (see Supplementary Figure 5, \cite{suppinfo}). In fact, when normalizing networks by initial values $L_c(0)$ and $\lambda_2(0)$, pruned networks appear to fall into two classes of either belonging to large cycles, high connectivity for relative resistance pruning $(T<1)$ or low connectivity, locally tree-like for random pruning $(T\geq1)$ independent of the underlying initial network architecture (Fig.~\ref{fig4}C). The two classes of network architecture suggest that the ideal scheme of maximum relative resistance pruning is not needed but indeed only the distinction of small temperature in Boltzmann sampling, i.e. $T<1$, is a necessary condition to form efficient networks. 

Finally, we revisit our initial prediction that in planar networks a vertex degree of 3 and marginally a vertex degree of 4 satisfy network efficiency, i.e.~$\rho_{ij}\leq1$ by quantifying the evolution of the node degree distribution for the random Erdős–Rényi networks starting out from a large average node degree ($\langle \nu \rangle \simeq 15$). As edges are removed, the node degree decay is identical for any pruning scheme, since it is only controlled by the number of edges, reduced by one at each step. However, the final distributions of node degrees are quite different: random pruning produces broad node-degree distributions ($\langle \nu\rangle\simeq 2.9 \pm 1.6$, signal to noise 1.8), whereas maximum relative resistance pruning converges toward a narrow distribution centered around $\langle \nu \rangle\simeq3.5 \pm 0.8$, signal to noise of 4 (Fig.~\ref{fig4}D). Thus, pruning edges based on relative resistance homogenizes the network topology to a node degree just below the predicted margin of node degree 4.


In this work, we show that adapting transport networks by pruning superfluous edges based on their relative resistance yields efficient networks -- using only flow magnitude as edge-based readout, without the need for a global cost function or specifying boundary conditions. Compared to alternative pruning schemes, networks adapted by relative resistance result in lower dissipation, are well-connected, loop-rich, and homogeneous in node degree.

Pruning by relative resistance is inspired by experimental observations in a living, continuously remodeling transport network, the vein networks of {\it P.~polycephalum} \cite{Marbach.2023}. And yet, pruning networks is not limited to transport networks at all, but resembles the graph theoretic challenge of sparsifying a graph but retaining global network properties by down sampling edges. Drawing the analogy to the classical Spielman-Srivastava algorithm sampling edges by the effective resistance \cite{Spielman2011a} suggests sampling edges by $p_{ij}=1/(1+\rho_{ij})$ as a biologically inspired approach worth future investigation in the graph theoretic context.

Compared to biological inspiration we here, for simplicity, only considered pruning of edges without redistribution of the freed material resource among neighboring edges, unlike adaptation in biological systems, where, for instance, cells of regressing blood vessel move to neighboring vessels \cite{Ricard2015}. Thus, incorporating conductance redistribution under a fixed building cost would provide an even closer connection to existing adaptive flow models based on global dissipation minimization \cite{Corson2010}. 

We find that Boltzmann-like weighting by relative resistance with $T<1$ is sufficient to obtain efficient networks, and yet biological network remodeling might well leverage the fact that the temperature-like parameter $T$ sets the balance between exploiting lower dissipation and exploring alternative configurations, akin to simulated annealing. Experimentally our Boltzmann temperature $T$ could be inferred directly from the observed order of edge pruning, offering a quantitative measure of the degree to which biological networks' dynamics follow local optimization. More broadly, casting network remodeling in terms of a local, boundary-condition-free quantity offers a route to connect graph-theoretic notions of sparsification with the physics of self-organized, adaptive transport networks.

\begin{acknowledgments}
{\it Acknowledgments} — This work was supported by the European Research Council (ERC) under the European Union’s Horizon 2020 research and innovation program (grant agreement 947630, FlowMem) and the Human Frontier Science Program Organization through Research Grant number RGP0001/2021.

\end{acknowledgments}

\bibliography{bibliography}

\begin{thebibliography}{39}%
\makeatletter
\providecommand \@ifxundefined [1]{%
 \@ifx{#1\undefined}
}%
\providecommand \@ifnum [1]{%
 \ifnum #1\expandafter \@firstoftwo
 \else \expandafter \@secondoftwo
 \fi
}%
\providecommand \@ifx [1]{%
 \ifx #1\expandafter \@firstoftwo
 \else \expandafter \@secondoftwo
 \fi
}%
\providecommand \natexlab [1]{#1}%
\providecommand \enquote  [1]{``#1''}%
\providecommand \bibnamefont  [1]{#1}%
\providecommand \bibfnamefont [1]{#1}%
\providecommand \citenamefont [1]{#1}%
\providecommand \href@noop [0]{\@secondoftwo}%
\providecommand \href [0]{\begingroup \@sanitize@url \@href}%
\providecommand \@href[1]{\@@startlink{#1}\@@href}%
\providecommand \@@href[1]{\endgroup#1\@@endlink}%
\providecommand \@sanitize@url [0]{\catcode `\\12\catcode `\$12\catcode
  `\&12\catcode `\#12\catcode `\^12\catcode `\_12\catcode `\%12\relax}%
\providecommand \@@startlink[1]{}%
\providecommand \@@endlink[0]{}%
\providecommand \url  [0]{\begingroup\@sanitize@url \@url }%
\providecommand \@url [1]{\endgroup\@href {#1}{\urlprefix }}%
\providecommand \urlprefix  [0]{URL }%
\providecommand \Eprint [0]{\href }%
\providecommand \doibase [0]{https://doi.org/}%
\providecommand \selectlanguage [0]{\@gobble}%
\providecommand \bibinfo  [0]{\@secondoftwo}%
\providecommand \bibfield  [0]{\@secondoftwo}%
\providecommand \translation [1]{[#1]}%
\providecommand \BibitemOpen [0]{}%
\providecommand \bibitemStop [0]{}%
\providecommand \bibitemNoStop [0]{.\EOS\space}%
\providecommand \EOS [0]{\spacefactor3000\relax}%
\providecommand \BibitemShut  [1]{\csname bibitem#1\endcsname}%
\let\auto@bib@innerbib\@empty
\bibitem [{\citenamefont {Spielman}\ and\ \citenamefont
  {Srivastava}(2011)}]{Spielman2011a}%
  \BibitemOpen
  \bibfield  {author} {\bibinfo {author} {\bibfnamefont {D.~A.}\ \bibnamefont
  {Spielman}}\ and\ \bibinfo {author} {\bibfnamefont {N.}~\bibnamefont
  {Srivastava}},\ }\bibfield  {title} {\bibinfo {title} {{Graph Sparsification
  by Effective Resistances}},\ }\href {https://doi.org/10.1137/080734029}
  {\bibfield  {journal} {\bibinfo  {journal} {SIAM Journal on Computing}\
  }\textbf {\bibinfo {volume} {40}},\ \bibinfo {pages} {1913} (\bibinfo {year}
  {2011})}\BibitemShut {NoStop}%
\bibitem [{\citenamefont {Forrow}\ \emph {et~al.}(2018)\citenamefont {Forrow},
  \citenamefont {Woodhouse},\ and\ \citenamefont {Dunkel}}]{Forrow2018}%
  \BibitemOpen
  \bibfield  {author} {\bibinfo {author} {\bibfnamefont {A.}~\bibnamefont
  {Forrow}}, \bibinfo {author} {\bibfnamefont {F.~G.}\ \bibnamefont
  {Woodhouse}},\ and\ \bibinfo {author} {\bibfnamefont {J.}~\bibnamefont
  {Dunkel}},\ }\bibfield  {title} {\bibinfo {title} {{Functional Control of
  Network Dynamics Using Designed Laplacian Spectra}},\ }\href
  {https://doi.org/10.1103/PhysRevX.8.041043} {\bibfield  {journal} {\bibinfo
  {journal} {Physical Review X}\ }\textbf {\bibinfo {volume} {8}},\ \bibinfo
  {pages} {41043} (\bibinfo {year} {2018})}\BibitemShut {NoStop}%
\bibitem [{\citenamefont {Kaiser}\ \emph {et~al.}(2020)\citenamefont {Kaiser},
  \citenamefont {Ronellenfitsch},\ and\ \citenamefont {Witthaut}}]{Kaiser2020}%
  \BibitemOpen
  \bibfield  {author} {\bibinfo {author} {\bibfnamefont {F.}~\bibnamefont
  {Kaiser}}, \bibinfo {author} {\bibfnamefont {H.}~\bibnamefont
  {Ronellenfitsch}},\ and\ \bibinfo {author} {\bibfnamefont {D.}~\bibnamefont
  {Witthaut}},\ }\bibfield  {title} {\bibinfo {title} {{Discontinuous
  transition to loop formation in optimal supply networks}},\ }\href
  {https://doi.org/10.1038/s41467-020-19567-2} {\bibfield  {journal} {\bibinfo
  {journal} {Nature Communications}\ }\textbf {\bibinfo {volume} {11}},\
  \bibinfo {pages} {5796} (\bibinfo {year} {2020})}\BibitemShut {NoStop}%
\bibitem [{\citenamefont {Bohn}\ and\ \citenamefont
  {Magnasco}(2007)}]{Bohn2007}%
  \BibitemOpen
  \bibfield  {author} {\bibinfo {author} {\bibfnamefont {S.}~\bibnamefont
  {Bohn}}\ and\ \bibinfo {author} {\bibfnamefont {M.~O.}\ \bibnamefont
  {Magnasco}},\ }\bibfield  {title} {\bibinfo {title} {{Structure, scaling, and
  phase transition in the optimal transport network}},\ }\href
  {https://doi.org/10.1103/PhysRevLett.98.088702} {\bibfield  {journal}
  {\bibinfo  {journal} {Physical Review Letters}\ }\textbf {\bibinfo {volume}
  {98}},\ \bibinfo {pages} {3} (\bibinfo {year} {2007})}\BibitemShut {NoStop}%
\bibitem [{\citenamefont {Corson}(2010)}]{Corson2010}%
  \BibitemOpen
  \bibfield  {author} {\bibinfo {author} {\bibfnamefont {F.}~\bibnamefont
  {Corson}},\ }\bibfield  {title} {\bibinfo {title} {{Fluctuations and
  Redundancy in Optimal Transport Networks}},\ }\href
  {https://doi.org/10.1103/PhysRevLett.104.048703} {\bibfield  {journal}
  {\bibinfo  {journal} {Physical Review Letters}\ }\textbf {\bibinfo {volume}
  {104}},\ \bibinfo {pages} {048703} (\bibinfo {year} {2010})}\BibitemShut
  {NoStop}%
\bibitem [{\citenamefont {Katifori}\ \emph {et~al.}(2010)\citenamefont
  {Katifori}, \citenamefont {Sz{\"{o}}llősi},\ and\ \citenamefont
  {Magnasco}}]{Katifori2010}%
  \BibitemOpen
  \bibfield  {author} {\bibinfo {author} {\bibfnamefont {E.}~\bibnamefont
  {Katifori}}, \bibinfo {author} {\bibfnamefont {G.~J.}\ \bibnamefont
  {Sz{\"{o}}llősi}},\ and\ \bibinfo {author} {\bibfnamefont {M.~O.}\
  \bibnamefont {Magnasco}},\ }\bibfield  {title} {\bibinfo {title} {{Damage and
  Fluctuations Induce Loops in Optimal Transport Networks}},\ }\href
  {https://doi.org/10.1103/PhysRevLett.104.048704} {\bibfield  {journal}
  {\bibinfo  {journal} {Physical Review Letters}\ }\textbf {\bibinfo {volume}
  {104}},\ \bibinfo {pages} {048704} (\bibinfo {year} {2010})}\BibitemShut
  {NoStop}%
\bibitem [{\citenamefont {Cai}\ \emph {et~al.}(2023)\citenamefont {Cai},
  \citenamefont {Chen},\ and\ \citenamefont {Peng}}]{Cai2023a}%
  \BibitemOpen
  \bibfield  {author} {\bibinfo {author} {\bibfnamefont {D.}~\bibnamefont
  {Cai}}, \bibinfo {author} {\bibfnamefont {X.}~\bibnamefont {Chen}},\ and\
  \bibinfo {author} {\bibfnamefont {P.}~\bibnamefont {Peng}},\ }\href
  {https://doi.org/10.4230/LIPIcs.ESA.2023.29} {\bibinfo {title} {{Effective
  Resistances in Non-Expander Graphs}}} (\bibinfo {year} {2023}),\ \Eprint
  {https://arxiv.org/abs/2307.01218} {2307.01218} \BibitemShut {NoStop}%
\bibitem [{\citenamefont {Kirkegaard}\ and\ \citenamefont
  {Sneppen}(2020)}]{Kirkegaard2020}%
  \BibitemOpen
  \bibfield  {author} {\bibinfo {author} {\bibfnamefont {J.~B.}\ \bibnamefont
  {Kirkegaard}}\ and\ \bibinfo {author} {\bibfnamefont {K.}~\bibnamefont
  {Sneppen}},\ }\bibfield  {title} {\bibinfo {title} {{Optimal Transport Flows
  for Distributed Production Networks}},\ }\href
  {https://doi.org/10.1103/PhysRevLett.124.208101} {\bibfield  {journal}
  {\bibinfo  {journal} {Physical Review Letters}\ }\textbf {\bibinfo {volume}
  {124}},\ \bibinfo {pages} {208101} (\bibinfo {year} {2020})}\BibitemShut
  {NoStop}%
\bibitem [{\citenamefont {Dahl-Jensen}\ \emph {et~al.}(2018)\citenamefont
  {Dahl-Jensen}, \citenamefont {Yennek}, \citenamefont {Flasse}, \citenamefont
  {Larsen}, \citenamefont {Sever}, \citenamefont {Karremore}, \citenamefont
  {Novak}, \citenamefont {Sneppen},\ and\ \citenamefont
  {Grapin-Botton}}]{DahlJensen2018}%
  \BibitemOpen
  \bibfield  {author} {\bibinfo {author} {\bibfnamefont {S.~B.}\ \bibnamefont
  {Dahl-Jensen}}, \bibinfo {author} {\bibfnamefont {S.}~\bibnamefont {Yennek}},
  \bibinfo {author} {\bibfnamefont {L.}~\bibnamefont {Flasse}}, \bibinfo
  {author} {\bibfnamefont {H.~L.}\ \bibnamefont {Larsen}}, \bibinfo {author}
  {\bibfnamefont {D.}~\bibnamefont {Sever}}, \bibinfo {author} {\bibfnamefont
  {G.}~\bibnamefont {Karremore}}, \bibinfo {author} {\bibfnamefont
  {I.}~\bibnamefont {Novak}}, \bibinfo {author} {\bibfnamefont
  {K.}~\bibnamefont {Sneppen}},\ and\ \bibinfo {author} {\bibfnamefont
  {A.}~\bibnamefont {Grapin-Botton}},\ }\bibfield  {title} {\bibinfo {title}
  {{Deconstructing the principles of ductal network formation in the
  pancreas}},\ }\href {https://doi.org/10.1371/journal.pbio.2002842} {\bibfield
   {journal} {\bibinfo  {journal} {PLoS Biology}\ }\textbf {\bibinfo {volume}
  {16}},\ \bibinfo {pages} {1} (\bibinfo {year} {2018})}\BibitemShut {NoStop}%
\bibitem [{\citenamefont {Katifori}(2018)}]{Katifori2018}%
  \BibitemOpen
  \bibfield  {author} {\bibinfo {author} {\bibfnamefont {E.}~\bibnamefont
  {Katifori}},\ }\bibfield  {title} {\bibinfo {title} {{The transport network
  of a leaf}},\ }\href {https://doi.org/10.1016/j.crhy.2018.10.007} {\bibfield
  {journal} {\bibinfo  {journal} {Comptes Rendus Physique}\ }\textbf {\bibinfo
  {volume} {19}},\ \bibinfo {pages} {244} (\bibinfo {year} {2018})}\BibitemShut
  {NoStop}%
\bibitem [{\citenamefont {Fricker}\ \emph {et~al.}(2017)\citenamefont
  {Fricker}, \citenamefont {Heaton}, \citenamefont {Jones},\ and\ \citenamefont
  {Boddy}}]{Fricker2017a}%
  \BibitemOpen
  \bibfield  {author} {\bibinfo {author} {\bibfnamefont {M.~D.}\ \bibnamefont
  {Fricker}}, \bibinfo {author} {\bibfnamefont {L.~L.~M.}\ \bibnamefont
  {Heaton}}, \bibinfo {author} {\bibfnamefont {N.~S.}\ \bibnamefont {Jones}},\
  and\ \bibinfo {author} {\bibfnamefont {L.}~\bibnamefont {Boddy}},\ }\bibfield
   {title} {\bibinfo {title} {{The Mycelium as a Network}},\ }in\ \href
  {https://doi.org/10.1128/9781555819583.ch15} {\emph {\bibinfo {booktitle}
  {The Fungal Kingdom}}},\ \bibinfo {series and number} {\bibinfo {number}
  {September}}\ (\bibinfo  {publisher} {ASM Press},\ \bibinfo {address}
  {Washington, DC, USA},\ \bibinfo {year} {2017})\ pp.\ \bibinfo {pages}
  {335--367}\BibitemShut {NoStop}%
\bibitem [{\citenamefont {Tero}\ \emph {et~al.}(2010)\citenamefont {Tero},
  \citenamefont {Takagi}, \citenamefont {Saigusa}, \citenamefont {Ito},
  \citenamefont {Bebber}, \citenamefont {Fricker}, \citenamefont {Yumiki},
  \citenamefont {Kobayashi},\ and\ \citenamefont {Nakagaki}}]{Tero2010}%
  \BibitemOpen
  \bibfield  {author} {\bibinfo {author} {\bibfnamefont {A.}~\bibnamefont
  {Tero}}, \bibinfo {author} {\bibfnamefont {S.}~\bibnamefont {Takagi}},
  \bibinfo {author} {\bibfnamefont {T.}~\bibnamefont {Saigusa}}, \bibinfo
  {author} {\bibfnamefont {K.}~\bibnamefont {Ito}}, \bibinfo {author}
  {\bibfnamefont {D.~P.}\ \bibnamefont {Bebber}}, \bibinfo {author}
  {\bibfnamefont {M.~D.}\ \bibnamefont {Fricker}}, \bibinfo {author}
  {\bibfnamefont {K.}~\bibnamefont {Yumiki}}, \bibinfo {author} {\bibfnamefont
  {R.}~\bibnamefont {Kobayashi}},\ and\ \bibinfo {author} {\bibfnamefont
  {T.}~\bibnamefont {Nakagaki}},\ }\bibfield  {title} {\bibinfo {title} {{Rules
  for Biologically Inspired Adaptive Network Design}},\ }\href
  {https://doi.org/10.1126/science.1177894} {\bibfield  {journal} {\bibinfo
  {journal} {Science}\ }\textbf {\bibinfo {volume} {327}},\ \bibinfo {pages}
  {439} (\bibinfo {year} {2010})}\BibitemShut {NoStop}%
\bibitem [{\citenamefont {Alim}(2018)}]{Alim2018}%
  \BibitemOpen
  \bibfield  {author} {\bibinfo {author} {\bibfnamefont {K.}~\bibnamefont
  {Alim}},\ }\bibfield  {title} {\bibinfo {title} {{Fluid flows shaping
  organism morphology}},\ }\href {https://doi.org/10.1098/rstb.2017.0112}
  {\bibfield  {journal} {\bibinfo  {journal} {Philosophical Transactions of the
  Royal Society B: Biological Sciences}\ }\textbf {\bibinfo {volume} {373}},\
  \bibinfo {pages} {20170112} (\bibinfo {year} {2018})}\BibitemShut {NoStop}%
\bibitem [{\citenamefont {Murray}(1926)}]{Murray1926}%
  \BibitemOpen
  \bibfield  {author} {\bibinfo {author} {\bibfnamefont {C.~D.}\ \bibnamefont
  {Murray}},\ }\bibfield  {title} {\bibinfo {title} {{The Physiological
  Principle of Minimum Work: II. Oxygen Exchange in Capillaries}},\ }\href
  {https://doi.org/10.1073/pnas.12.5.299} {\bibfield  {journal} {\bibinfo
  {journal} {Proceedings of the National Academy of Sciences}\ }\textbf
  {\bibinfo {volume} {12}},\ \bibinfo {pages} {299} (\bibinfo {year}
  {1926})}\BibitemShut {NoStop}%
\bibitem [{\citenamefont {Durand}(2007)}]{Durand2007}%
  \BibitemOpen
  \bibfield  {author} {\bibinfo {author} {\bibfnamefont {M.}~\bibnamefont
  {Durand}},\ }\bibfield  {title} {\bibinfo {title} {{Structure of optimal
  transport networks subject to a global constraint}},\ }\href
  {https://doi.org/10.1103/PhysRevLett.98.088701} {\bibfield  {journal}
  {\bibinfo  {journal} {Physical Review Letters}\ }\textbf {\bibinfo {volume}
  {98}},\ \bibinfo {pages} {1} (\bibinfo {year} {2007})}\BibitemShut {NoStop}%
\bibitem [{\citenamefont {Hu}\ and\ \citenamefont {Cai}(2013)}]{Hu2013}%
  \BibitemOpen
  \bibfield  {author} {\bibinfo {author} {\bibfnamefont {D.}~\bibnamefont
  {Hu}}\ and\ \bibinfo {author} {\bibfnamefont {D.}~\bibnamefont {Cai}},\
  }\bibfield  {title} {\bibinfo {title} {{Adaptation and Optimization of
  Biological Transport Networks}},\ }\href
  {https://doi.org/10.1103/PhysRevLett.111.138701} {\bibfield  {journal}
  {\bibinfo  {journal} {Physical Review Letters}\ }\textbf {\bibinfo {volume}
  {111}},\ \bibinfo {pages} {138701} (\bibinfo {year} {2013})}\BibitemShut
  {NoStop}%
\bibitem [{\citenamefont {Ronellenfitsch}\ and\ \citenamefont
  {Katifori}(2016)}]{Ronellenfitsch2016}%
  \BibitemOpen
  \bibfield  {author} {\bibinfo {author} {\bibfnamefont {H.}~\bibnamefont
  {Ronellenfitsch}}\ and\ \bibinfo {author} {\bibfnamefont {E.}~\bibnamefont
  {Katifori}},\ }\bibfield  {title} {\bibinfo {title} {{Global Optimization,
  Local Adaptation, and the Role of Growth in Distribution Networks}},\ }\href
  {https://doi.org/10.1103/PhysRevLett.117.138301} {\bibfield  {journal}
  {\bibinfo  {journal} {Physical Review Letters}\ }\textbf {\bibinfo {volume}
  {117}},\ \bibinfo {pages} {138301} (\bibinfo {year} {2016})}\BibitemShut
  {NoStop}%
\bibitem [{\citenamefont {Ronellenfitsch}\ and\ \citenamefont
  {Katifori}(2019)}]{Ronellenfitsch.20190t}%
  \BibitemOpen
  \bibfield  {author} {\bibinfo {author} {\bibfnamefont {H.}~\bibnamefont
  {Ronellenfitsch}}\ and\ \bibinfo {author} {\bibfnamefont {E.}~\bibnamefont
  {Katifori}},\ }\bibfield  {title} {\bibinfo {title} {{Phenotypes of vascular
  flow networks}},\ }\href {https://doi.org/10.1103/physrevlett.123.248101}
  {\bibfield  {journal} {\bibinfo  {journal} {Physical Review Letters}\
  }\textbf {\bibinfo {volume} {123}},\ \bibinfo {pages} {248101} (\bibinfo
  {year} {2019})}\BibitemShut {NoStop}%
\bibitem [{\citenamefont {Meigel}\ and\ \citenamefont
  {Alim}(2018)}]{Meigel2018a}%
  \BibitemOpen
  \bibfield  {author} {\bibinfo {author} {\bibfnamefont {F.~J.}\ \bibnamefont
  {Meigel}}\ and\ \bibinfo {author} {\bibfnamefont {K.}~\bibnamefont {Alim}},\
  }\bibfield  {title} {\bibinfo {title} {{Flow rate of transport network
  controls uniform metabolite supply to tissue}},\ }\bibfield  {journal}
  {\bibinfo  {journal} {Journal of the Royal Society Interface}\ }\textbf
  {\bibinfo {volume} {15}},\ \href {https://doi.org/10.1098/rsif.2018.0075}
  {10.1098/rsif.2018.0075} (\bibinfo {year} {2018})\BibitemShut {NoStop}%
\bibitem [{\citenamefont {Kramer}\ and\ \citenamefont
  {Modes}(2023)}]{Kramer2023}%
  \BibitemOpen
  \bibfield  {author} {\bibinfo {author} {\bibfnamefont {F.}~\bibnamefont
  {Kramer}}\ and\ \bibinfo {author} {\bibfnamefont {C.~D.}\ \bibnamefont
  {Modes}},\ }\bibfield  {title} {\bibinfo {title} {{Biological flow networks:
  Antagonism between hydrodynamic and metabolic stimuli as driver of
  topological transitions}},\ }\href
  {https://doi.org/10.1103/PhysRevResearch.5.023106} {\bibfield  {journal}
  {\bibinfo  {journal} {Physical Review Research}\ }\textbf {\bibinfo {volume}
  {5}},\ \bibinfo {pages} {1} (\bibinfo {year} {2023})}\BibitemShut {NoStop}%
\bibitem [{\citenamefont {Zareei}\ \emph {et~al.}(2022)\citenamefont {Zareei},
  \citenamefont {Pan},\ and\ \citenamefont {Amir}}]{Zareei2022}%
  \BibitemOpen
  \bibfield  {author} {\bibinfo {author} {\bibfnamefont {A.}~\bibnamefont
  {Zareei}}, \bibinfo {author} {\bibfnamefont {D.}~\bibnamefont {Pan}},\ and\
  \bibinfo {author} {\bibfnamefont {A.}~\bibnamefont {Amir}},\ }\bibfield
  {title} {\bibinfo {title} {{Temporal Evolution of Erosion in Pore Networks:
  From Homogenization to Instability - Supplementary Informations}},\ }\href
  {https://doi.org/10.1103/PhysRevLett.128.234501} {\bibfield  {journal}
  {\bibinfo  {journal} {Physical Review Letters}\ }\textbf {\bibinfo {volume}
  {128}},\ \bibinfo {pages} {1} (\bibinfo {year} {2022})}\BibitemShut {NoStop}%
\bibitem [{\citenamefont {Bouvard}\ \emph {et~al.}(2025)\citenamefont
  {Bouvard}, \citenamefont {Basu}, \citenamefont {Leu}, \citenamefont {Bektas},
  \citenamefont {R{\"{a}}dler}, \citenamefont {Amselem},\ and\ \citenamefont
  {Alim}}]{Bouvard2025}%
  \BibitemOpen
  \bibfield  {author} {\bibinfo {author} {\bibfnamefont {J.}~\bibnamefont
  {Bouvard}}, \bibinfo {author} {\bibfnamefont {S.}~\bibnamefont {Basu}},
  \bibinfo {author} {\bibfnamefont {C.}~\bibnamefont {Leu}}, \bibinfo {author}
  {\bibfnamefont {O.}~\bibnamefont {Bektas}}, \bibinfo {author} {\bibfnamefont
  {J.~O.}\ \bibnamefont {R{\"{a}}dler}}, \bibinfo {author} {\bibfnamefont
  {G.}~\bibnamefont {Amselem}},\ and\ \bibinfo {author} {\bibfnamefont
  {K.}~\bibnamefont {Alim}},\ }\bibfield  {title} {\bibinfo {title}
  {{Self-Organized Homogenization of Flow Networks}},\ }\href
  {https://doi.org/10.1103/j5ch-4vkh} {\bibfield  {journal} {\bibinfo
  {journal} {Physical Review X}\ }\textbf {\bibinfo {volume} {15}},\ \bibinfo
  {pages} {041038} (\bibinfo {year} {2025})}\BibitemShut {NoStop}%
\bibitem [{\citenamefont {Bhattacharyya}\ \emph {et~al.}(2022)\citenamefont
  {Bhattacharyya}, \citenamefont {Zwicker},\ and\ \citenamefont
  {Alim}}]{Bhattacharyya.2022}%
  \BibitemOpen
  \bibfield  {author} {\bibinfo {author} {\bibfnamefont {K.}~\bibnamefont
  {Bhattacharyya}}, \bibinfo {author} {\bibfnamefont {D.}~\bibnamefont
  {Zwicker}},\ and\ \bibinfo {author} {\bibfnamefont {K.}~\bibnamefont
  {Alim}},\ }\bibfield  {title} {\bibinfo {title} {{Memory formation in
  adaptive networks}},\ }\href {https://doi.org/10.1103/physrevlett.129.028101}
  {\bibfield  {journal} {\bibinfo  {journal} {Physical Review Letters}\
  }\textbf {\bibinfo {volume} {129}},\ \bibinfo {pages} {028101} (\bibinfo
  {year} {2022})}\BibitemShut {NoStop}%
\bibitem [{\citenamefont {Stern}\ \emph {et~al.}(2025)\citenamefont {Stern},
  \citenamefont {Guzman}, \citenamefont {Martins}, \citenamefont {Liu},\ and\
  \citenamefont {Balasubramanian}}]{Stern2025}%
  \BibitemOpen
  \bibfield  {author} {\bibinfo {author} {\bibfnamefont {M.}~\bibnamefont
  {Stern}}, \bibinfo {author} {\bibfnamefont {M.}~\bibnamefont {Guzman}},
  \bibinfo {author} {\bibfnamefont {F.}~\bibnamefont {Martins}}, \bibinfo
  {author} {\bibfnamefont {A.~J.}\ \bibnamefont {Liu}},\ and\ \bibinfo {author}
  {\bibfnamefont {V.}~\bibnamefont {Balasubramanian}},\ }\bibfield  {title}
  {\bibinfo {title} {{Physical Networks Become What They Learn}},\ }\href
  {https://doi.org/10.1103/PhysRevLett.134.147402} {\bibfield  {journal}
  {\bibinfo  {journal} {Physical Review Letters}\ }\textbf {\bibinfo {volume}
  {134}},\ \bibinfo {pages} {147402} (\bibinfo {year} {2025})}\BibitemShut
  {NoStop}%
\bibitem [{\citenamefont {Anisetti}\ \emph {et~al.}(2023)\citenamefont
  {Anisetti}, \citenamefont {Scellier},\ and\ \citenamefont
  {Schwarz}}]{Anisetti2023}%
  \BibitemOpen
  \bibfield  {author} {\bibinfo {author} {\bibfnamefont {V.~R.}\ \bibnamefont
  {Anisetti}}, \bibinfo {author} {\bibfnamefont {B.}~\bibnamefont {Scellier}},\
  and\ \bibinfo {author} {\bibfnamefont {J.~M.}\ \bibnamefont {Schwarz}},\
  }\bibfield  {title} {\bibinfo {title} {{Learning by non-interfering feedback
  chemical signaling in physical networks}},\ }\href
  {https://doi.org/10.1103/PhysRevResearch.5.023024} {\bibfield  {journal}
  {\bibinfo  {journal} {Physical Review Research}\ }\textbf {\bibinfo {volume}
  {5}},\ \bibinfo {pages} {023024} (\bibinfo {year} {2023})}\BibitemShut
  {NoStop}%
\bibitem [{\citenamefont {Gavrilchenko}\ and\ \citenamefont
  {Katifori}(2019)}]{Gavrilchenko2019}%
  \BibitemOpen
  \bibfield  {author} {\bibinfo {author} {\bibfnamefont {T.}~\bibnamefont
  {Gavrilchenko}}\ and\ \bibinfo {author} {\bibfnamefont {E.}~\bibnamefont
  {Katifori}},\ }\bibfield  {title} {\bibinfo {title} {{Resilience in
  hierarchical fluid flow networks}},\ }\href
  {https://doi.org/10.1103/PhysRevE.99.012321} {\bibfield  {journal} {\bibinfo
  {journal} {Physical Review E}\ }\textbf {\bibinfo {volume} {99}},\ \bibinfo
  {pages} {012321} (\bibinfo {year} {2019})}\BibitemShut {NoStop}%
\bibitem [{\citenamefont {Marbach}\ \emph
  {et~al.}(2023{\natexlab{a}})\citenamefont {Marbach}, \citenamefont {Ziethen},
  \citenamefont {Bastin}, \citenamefont {Bäuerle},\ and\ \citenamefont
  {Alim}}]{Marbach.2023}%
  \BibitemOpen
  \bibfield  {author} {\bibinfo {author} {\bibfnamefont {S.}~\bibnamefont
  {Marbach}}, \bibinfo {author} {\bibfnamefont {N.}~\bibnamefont {Ziethen}},
  \bibinfo {author} {\bibfnamefont {L.}~\bibnamefont {Bastin}}, \bibinfo
  {author} {\bibfnamefont {F.~K.}\ \bibnamefont {Bäuerle}},\ and\ \bibinfo
  {author} {\bibfnamefont {K.}~\bibnamefont {Alim}},\ }\bibfield  {title}
  {\bibinfo {title} {{Vein fate determined by flow-based but time-delayed
  integration of network architecture}},\ }\href
  {https://doi.org/10.7554/elife.78100} {\bibfield  {journal} {\bibinfo
  {journal} {eLife}\ }\textbf {\bibinfo {volume} {12}},\ \bibinfo {pages}
  {e78100} (\bibinfo {year} {2023}{\natexlab{a}})}\BibitemShut {NoStop}%
\bibitem [{\citenamefont {{Le Verge-Serandour}}\ and\ \citenamefont
  {Alim}(2024)}]{LeVergeSerandour2024}%
  \BibitemOpen
  \bibfield  {author} {\bibinfo {author} {\bibfnamefont {M.}~\bibnamefont {{Le
  Verge-Serandour}}}\ and\ \bibinfo {author} {\bibfnamefont {K.}~\bibnamefont
  {Alim}},\ }\bibfield  {title} {\bibinfo {title} {{Physarum polycephalum :
  Smart Network Adaptation}},\ }\href
  {https://doi.org/10.1146/annurev-conmatphys-040821-115312} {\bibfield
  {journal} {\bibinfo  {journal} {Annual Review of Condensed Matter Physics}\
  }\textbf {\bibinfo {volume} {15}},\ \bibinfo {pages} {72} (\bibinfo {year}
  {2024})}\BibitemShut {NoStop}%
\bibitem [{\citenamefont {Marbach}\ \emph
  {et~al.}(2023{\natexlab{b}})\citenamefont {Marbach}, \citenamefont
  {Ziethen},\ and\ \citenamefont {Alim}}]{Marbach.2023rjf}%
  \BibitemOpen
  \bibfield  {author} {\bibinfo {author} {\bibfnamefont {S.}~\bibnamefont
  {Marbach}}, \bibinfo {author} {\bibfnamefont {N.}~\bibnamefont {Ziethen}},\
  and\ \bibinfo {author} {\bibfnamefont {K.}~\bibnamefont {Alim}},\ }\bibfield
  {title} {\bibinfo {title} {{Vascular adaptation model from force balance:
  Physarum polycephalum as a case study}},\ }\href
  {https://doi.org/10.1088/1367-2630/ad1488} {\bibfield  {journal} {\bibinfo
  {journal} {New Journal of Physics}\ }\textbf {\bibinfo {volume} {25}},\
  \bibinfo {pages} {123052} (\bibinfo {year} {2023}{\natexlab{b}})}\BibitemShut
  {NoStop}%
\bibitem [{\citenamefont {D{\"{o}}rfler}\ and\ \citenamefont
  {Bullo}(2013)}]{Dorfler2013}%
  \BibitemOpen
  \bibfield  {author} {\bibinfo {author} {\bibfnamefont {F.}~\bibnamefont
  {D{\"{o}}rfler}}\ and\ \bibinfo {author} {\bibfnamefont {F.}~\bibnamefont
  {Bullo}},\ }\bibfield  {title} {\bibinfo {title} {{Kron reduction of graphs
  with applications to electrical networks}},\ }\href
  {https://doi.org/10.1109/TCSI.2012.2215780} {\bibfield  {journal} {\bibinfo
  {journal} {IEEE Transactions on Circuits and Systems I: Regular Papers}\
  }\textbf {\bibinfo {volume} {60}},\ \bibinfo {pages} {150} (\bibinfo {year}
  {2013})}\BibitemShut {NoStop}%
\bibitem [{sup()}]{suppinfo}%
  \BibitemOpen
  \href@noop {} {}\bibinfo {note} {Supplemental Material is included in this
  version of the article.}\BibitemShut {Stop}%
\bibitem [{\citenamefont {Ghosh}\ \emph {et~al.}(2008)\citenamefont {Ghosh},
  \citenamefont {Boyd},\ and\ \citenamefont {Saberi}}]{Ghosh2008}%
  \BibitemOpen
  \bibfield  {author} {\bibinfo {author} {\bibfnamefont {A.}~\bibnamefont
  {Ghosh}}, \bibinfo {author} {\bibfnamefont {S.}~\bibnamefont {Boyd}},\ and\
  \bibinfo {author} {\bibfnamefont {A.}~\bibnamefont {Saberi}},\ }\bibfield
  {title} {\bibinfo {title} {{Minimizing Effective Resistance of a Graph}},\
  }\href {https://doi.org/10.1137/050645452} {\bibfield  {journal} {\bibinfo
  {journal} {SIAM Review}\ }\textbf {\bibinfo {volume} {50}},\ \bibinfo {pages}
  {37} (\bibinfo {year} {2008})}\BibitemShut {NoStop}%
\bibitem [{\citenamefont {Tetali}(1991)}]{Tetali1991}%
  \BibitemOpen
  \bibfield  {author} {\bibinfo {author} {\bibfnamefont {P.}~\bibnamefont
  {Tetali}},\ }\bibfield  {title} {\bibinfo {title} {{Random walks and the
  effective resistance of networks}},\ }\href
  {https://doi.org/10.1007/BF01046996} {\bibfield  {journal} {\bibinfo
  {journal} {Journal of Theoretical Probability}\ }\textbf {\bibinfo {volume}
  {4}},\ \bibinfo {pages} {101} (\bibinfo {year} {1991})}\BibitemShut {NoStop}%
\bibitem [{\citenamefont {Klein}\ and\ \citenamefont
  {Randi{\'{c}}}(1993)}]{Klein1993}%
  \BibitemOpen
  \bibfield  {author} {\bibinfo {author} {\bibfnamefont {D.~J.}\ \bibnamefont
  {Klein}}\ and\ \bibinfo {author} {\bibfnamefont {M.}~\bibnamefont
  {Randi{\'{c}}}},\ }\bibfield  {title} {\bibinfo {title} {{Resistance
  distance}},\ }\href {https://doi.org/10.1007/BF01164627} {\bibfield
  {journal} {\bibinfo  {journal} {Journal of Mathematical Chemistry}\ }\textbf
  {\bibinfo {volume} {12}},\ \bibinfo {pages} {81} (\bibinfo {year}
  {1993})}\BibitemShut {NoStop}%
\bibitem [{\citenamefont {Manik}\ \emph {et~al.}(2017)\citenamefont {Manik},
  \citenamefont {Rohden}, \citenamefont {Ronellenfitsch}, \citenamefont
  {Zhang}, \citenamefont {Hallerberg}, \citenamefont {Witthaut},\ and\
  \citenamefont {Timme}}]{Manik2017}%
  \BibitemOpen
  \bibfield  {author} {\bibinfo {author} {\bibfnamefont {D.}~\bibnamefont
  {Manik}}, \bibinfo {author} {\bibfnamefont {M.}~\bibnamefont {Rohden}},
  \bibinfo {author} {\bibfnamefont {H.}~\bibnamefont {Ronellenfitsch}},
  \bibinfo {author} {\bibfnamefont {X.}~\bibnamefont {Zhang}}, \bibinfo
  {author} {\bibfnamefont {S.}~\bibnamefont {Hallerberg}}, \bibinfo {author}
  {\bibfnamefont {D.}~\bibnamefont {Witthaut}},\ and\ \bibinfo {author}
  {\bibfnamefont {M.}~\bibnamefont {Timme}},\ }\bibfield  {title} {\bibinfo
  {title} {{Network susceptibilities: Theory and applications}},\ }\href
  {https://doi.org/10.1103/PhysRevE.95.012319} {\bibfield  {journal} {\bibinfo
  {journal} {Physical Review E}\ }\textbf {\bibinfo {volume} {95}},\ \bibinfo
  {pages} {1} (\bibinfo {year} {2017})}\BibitemShut {NoStop}%
\bibitem [{\citenamefont {Fiedler}(1973)}]{Fiedler1973}%
  \BibitemOpen
  \bibfield  {author} {\bibinfo {author} {\bibfnamefont {M.}~\bibnamefont
  {Fiedler}},\ }\bibfield  {title} {\bibinfo {title} {{Algebraic connectivity
  of graphs}},\ }\href {https://doi.org/10.21136/CMJ.1973.101168} {\bibfield
  {journal} {\bibinfo  {journal} {Czechoslovak Mathematical Journal}\ }\textbf
  {\bibinfo {volume} {23}},\ \bibinfo {pages} {298} (\bibinfo {year}
  {1973})}\BibitemShut {NoStop}%
\bibitem [{\citenamefont {Ricard}\ and\ \citenamefont
  {Simons}(2015)}]{Ricard2015}%
  \BibitemOpen
  \bibfield  {author} {\bibinfo {author} {\bibfnamefont {N.}~\bibnamefont
  {Ricard}}\ and\ \bibinfo {author} {\bibfnamefont {M.}~\bibnamefont
  {Simons}},\ }\bibfield  {title} {\bibinfo {title} {{When It Is Better to
  Regress: Dynamics of Vascular Pruning}},\ }\href
  {https://doi.org/10.1371/journal.pbio.1002148} {\bibfield  {journal}
  {\bibinfo  {journal} {PLoS Biology}\ }\textbf {\bibinfo {volume} {13}},\
  \bibinfo {pages} {1} (\bibinfo {year} {2015})}\BibitemShut {NoStop}%
\bibitem [{\citenamefont {Bordenave}\ and\ \citenamefont
  {Caputo}(2015)}]{Bordenave2015}%
  \BibitemOpen
  \bibfield  {author} {\bibinfo {author} {\bibfnamefont {C.}~\bibnamefont
  {Bordenave}}\ and\ \bibinfo {author} {\bibfnamefont {P.}~\bibnamefont
  {Caputo}},\ }\bibfield  {title} {\bibinfo {title} {{Large deviations of
  empirical neighborhood distribution in sparse random graphs}},\ }\href
  {https://doi.org/10.1007/s00440-014-0590-8} {\bibfield  {journal} {\bibinfo
  {journal} {Probability Theory and Related Fields}\ }\textbf {\bibinfo
  {volume} {163}},\ \bibinfo {pages} {149} (\bibinfo {year}
  {2015})}\BibitemShut {NoStop}%
\bibitem [{\citenamefont {Luo}(2025)}]{Luo2025}%
  \BibitemOpen
  \bibfield  {author} {\bibinfo {author} {\bibfnamefont {K.}~\bibnamefont
  {Luo}},\ }\bibfield  {title} {\bibinfo {title} {{Locally Optimal Percolation
  for Network Resilience Dismantling via Fiedler Vector Gradient Iterative
  Attack}},\ }\href {https://doi.org/10.1063/5.0311738} {\bibfield  {journal}
  {\bibinfo  {journal} {arXiv}\ ,\ \bibinfo {pages} {1}} (\bibinfo {year}
  {2025})}\BibitemShut {NoStop}%
\end{thebibliography}%

\newpage
\clearpage
\onecolumngrid           
\begin{center}\textbf{\large Supplementary Information for \\Efficient Transport Networks through Edge Pruning by Relative Resistance}\end{center}
\setcounter{figure}{0}
\setcounter{table}{0}
\setcounter{equation}{0}
\setcounter{section}{0}
\renewcommand{\thefigure}{S\arabic{figure}}
\renewcommand{\theequation}{S\arabic{equation}}
\renewcommand{\thetable}{S\arabic{table}}

\section{Definitions and derivations}
\subsection{Relative resistance in planar regular networks}
\label{app:nodedegree}
A network $G = \lbrace V, E\rbrace$, with vertices $V$ and edges $E$ is represented by its graph Laplacian $L$,  a matrix with entries $L_{ij} = g_{ij} = -1/R_{ij}$ if $(i,j)\in E$, 0 else and, $L_{ii}=\sum_i 1/R_{ij}$, and $R_{ij}$ the resistance of edge $(i,j)$, $g_{ij}=1/R_{ij}$ being its conductance. Defining as $\vec{b}_{uv} = \vec{e}_u-\vec{e}_v$ a selector for nodes $u$ and $v$, we state the relationship between node pressures subsummed in the vector $\vec{p}$ and inflow at source node $u$ and outflow at sink node $v$ $\vec{b}_{uv}$ as $L\vec{p} = \vec{b}_{uv}$. Then, the pressure difference between the two nodes $u$ and $v$ that generates a flow of magnitude one, $p_u-p_v=R_{\rm{eff}}(u,v)$, defines the effective resistance $R_{\rm{eff}}(u,v)$: or equivalently $R_{\rm{eff}}(u,v) = \vec{b}_{uv}L^\dagger\vec{b}_{uv}$. Now considering source node $u$ and sink node $v$ as end nodes of edge $(u,v)$, we related relative resistance $\rho_{uv}$ and $R_{\rm{eff}}(u,v)$ by invoking Kirchhoff's law. The flow $Q_{\rm{in}}$ injected a $u$ splits into two parts: along edge $(u,v)$ $q_{uv}/Q_{\rm{in}}=(p_u-p_v)/R_{uv}$, and through the rest of the network $q_{\rm{net}}/Q_{\rm{in}}=(p_u-p_v)/R_{\rm{net}}(u,v)$ such that Kirchhoff's law asserts $1=(1/R_{uv}+1/R_{\rm{net}}(u,v))(p_u-p_v)$. Now, inserting the definition $R_{\rm{eff}}(u,v)=p_u-p_v$, and rearranging terms we find
\begin{equation}
\rho_{uv} = \frac{R_{uv}}{R_{\rm{eff}}(u,v)}-1.
\label{eqn_rr_reff}
\end{equation}
This relationship between the relative resistance and the effective resistance enables us to map to graph theoretic results. 

\begin{figure}[h]
    \centering
    \includegraphics[width=0.5\linewidth]{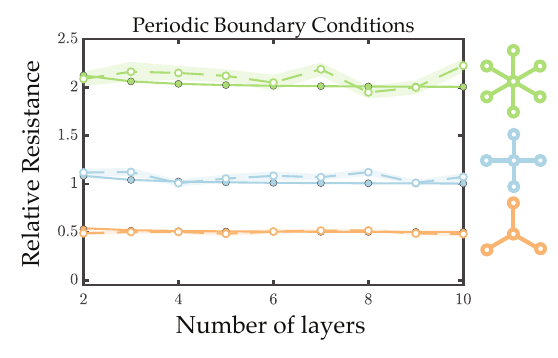}
    \caption{\textbf{Periodic boundary conditions do not change the relative resistance prediction.} Using triangular (green, $\nu=6$), square (blue, $\nu=4$), and hexagonal $\nu=3$ lattices, with periodic boundary conditions, we obtain the same numerical value for the central edge's relative resistance, matching our analytical prediction $\rho \sim \frac{\nu-2}{2}$. Full circles indicate uniform configurations ($R_{ij}=1$), inside white circles indicate noisy configurations ($R_{ij} =1\pm\varepsilon$, $\varepsilon = 0.2$, shaded regions the standard deviation, full line the analytical prediction. 100 simulations are ensemble-averaged for the noisy configurations.}
    \label{suppfig:regularnetworks_PBC}
\end{figure}

Now considering a regular network with $n$ nodes and node degree $\nu$, where all $m= \frac{\nu n}{2}$ edges are of uniform resistance $R_{ij}=1$ Foster theorem \cite{Ghosh2008} states $\sum_{u,v}g_{uv}R_{\rm{eff}}(u,v)= \rm{Tr}(LL^\dagger) = n-1$.  Since all edge resistances and the network are uniform, $R_{\rm{eff}}(u,v) = R_{\rm{eff}}$, the sum over edges simplifies to a product, $mR_{\rm{eff}} = n-1$. Now inserting the relationship between effective resistance and relative resistance Eq.~\eqref{eqn_rr_reff} and employing $m=\frac{\nu n}{2}$, we find
\begin{equation}
 \rho = \frac{\nu n}{2(n-1)}-1.   
\end{equation}
For large networks, i.e.~$n\to \infty$, $\rho\sim\frac{\nu-2}{2}$ in agreement with numerical results, see Figs. 1, \ref{suppfig:regularnetworks_PBC}. In the case where the resistances are not uniform, we have $\sum_{u,v}\frac{1}{1+\rho_{uv}} = n-1$.

\subsection{Change in relative resistance due to edge pruning}
\label{app:dipoledecay}
The relative resistance of an edge $(u,v)$ changes due to pruning an edge $(i,j)$ only because the effective resistance changes $\rho'(u,v) =\frac{R_{uv}}{R_{\rm{eff}}'(u,v)} - 1$. Now the change in effective resistance is $\delta R_{\rm{eff}}(u,v) =R'_{\rm{eff}}(u,v)-R_{\rm{eff}}(u,v)= b_{uv}^T \delta L^\dagger b_{uv}$ itself related to the change in in the pseudo inverse Laplacian. 
When removing edge $(i,j)$, the graph Laplacian $L$ changes to $L' = L - 1/R_{ij}b_{ij}\otimes b_{ij}^T$, where $\otimes$ denotes the outer product. Employing the Sherman-Morrison formula, we derive the change in the pseudo inverse Laplacian: $\left(A+u v^T\right)^{-1} = A^{-1} - \frac{A^{-1}uv^TA^{-1}}{1+v^TA^{-1}u}$, with now $A \rightarrow L$, $u\rightarrow - \sqrt{R_{ij}^{-1}}b_{ij}$, $v\rightarrow\sqrt{R_{ij}^{-1}}b_{ij}$, to be $(L')^\dagger = L^\dagger + R_{ij}^{-1} \frac{L^\dagger b_{ij}b_{ij}^TL^\dagger}{1-R_{ij}^{-1}b_{ij}^TL^\dagger b_{ij}}$. Now identifying $R_{\rm{eff}}(i,j)=b_{ij}L^\dagger b_{ij}$ and simplifying  the change in the pseudo inverse Laplacian is $\delta L^\dagger = R_{\rm{eff}}(i,j)/\rho_{ij} L^\dagger b_{ij}b_{ij}^TL^\dagger$. And thus the change in effective resistance follows to $\delta R_{\rm{eff}}(u,v) = \gamma_{ij}(b_{uv}^TL^\dagger b_{ij})^2$, where $\gamma_{ij} = \frac{g_{ij}}{1-g_{ij}R_{\rm{eff}}(i,j)} = \frac{1}{\rho_{ij}R_{\rm{eff}}(i,j)}$, using $R_{ij} = R_{\rm{eff}}(i,j)(1+\rho_{ij})$. Thus, the change in relative resistance $\delta \rho(u,v)=\rho'(u,v)-\rho(u,v)$ of edge $(u,v)$ due to pruning edge $(i,j)$ is 
\begin{equation}
\delta \rho(u,v)= -\frac{R_{uv}}{R_{\rm{eff}}(u,v)}\frac{\gamma_{ij}(b_{uv}^TL^\dagger b_{ij})^2}{R_{\rm{eff}}(u,v) + \gamma_{ij}(b_{uv}^TL^\dagger b_{ij})^2}.
\end{equation}
With this exact result at hand, we now turn to the continuum limit to derive scaling relationships.

Identifying in $c = b^T_{uv}L^\dagger b_{ij}$ the second derivative of a Green operator on a network ($G=L^\dagger$), we use that for planar networks in the continuum limit the Green operator scales with the graph distance $\ell_G$ between edges $(i,j)$ and $(u,v)$ as \cite{Manik2017}
$$G(\ell_G) = \begin{cases}\log(\ell_G), & d=2 \text{ (planar network)} \\ \ell_G^{2-d}, &d\neq 2.\end{cases}$$
As now $c\sim\nabla^2 G(\ell_G)$, we find $c\sim\ell_G^{-d}$  such that the change in relative resistance scales as
\begin{equation}
\delta \rho (u,v) \sim -\frac{\gamma_{ij} \ell_G^{-2d}}{R_{\rm{eff}(u,v)}+\gamma_{ij} \ell_G^{-2d}}=-\frac{1}{1+(\ell_G / \xi)^{2d}},
\end{equation}
with $\xi= (\gamma_{ij}/R_{\rm{eff}(u,v)})^{1/2d}$. Note that this scaling is only true as long as pruning induces a small change in the effective resistance $\gamma_{ij} c^2 \ll R_{\rm{eff}}(u,v)$, whereas for large changes the relative resistance saturates such that the relative resistance scales as $\delta \rho(u,v) \sim -\frac{R_{uv}}{R_{\rm{eff}}(u,v)}$.

Erdős–Rényi networks can be approximated, when the number of nodes is large, as a Galton-Watson tree with average number of neighbors $\langle \nu \rangle =np$ \cite{Bordenave2015}. Since the network is locally tree like, starting from a node, the number of nodes grows as $(\nu-1)^r$, for $r \ll \log n$. Thus, we consider a $\nu$-~regular tree for the derivation below, with $\nu>2$.
Taking two nodes $x$ and $y$ at a distance $r$, we want to find an expression for the associated Green's function $G(x,y)=G_r$, fulfilling $LG(x,y)=\delta(x,y)$. In matrix form this is equivalent to 
$$\sum_{z\in \partial_x}L_{xz}G(z,y)=\delta_{xy}$$
where $\partial_x$ is the set of neighboring vertices of $x$. For $x\neq y$, $\delta_{x,y}=0$, and for a $\nu$-regular tree, the Laplacian entries are $L_{xx}=\nu$ and $L_{xz}=-1$, thus $$\nu G(x,y)-\sum_{z\in\partial_x}G(z,y)=0.$$
Because the network is locally a tree, there exists only one path from $x$ to $y$, and only one node of $\partial_x$ is on this path, at a distance $r-1$ from $y$, while the $\nu-1$ other nodes are at a distance $r+1$ from $y$. Thus, using $G(x,y)=G_r$, we get
$$\nu G_r - G_{r-1} - (\nu-1)G_{r+1}=0.$$
Assuming the ansatz $G_r\sim q^r$ and substituting it in the recursive relation gives a second order polynomial in $q$, as $(\nu-1)q^2-\nu q+1 =0$, which admits two solutions: $q=1$ is a trivial solution and $q=\frac{1}{\nu-1}$ which gives
$$G_r \sim (\nu-1)^{-r}=e^{-r/\xi}, \quad \xi = \frac{1}{\ln(\nu-1)}.$$
with $\xi$ the correlation length. The scaling of $c^2=(\nabla^2G(\ell_G))^2\sim e^{-2\ell_G/\xi}$ follows, and thus, in the regime $\gamma_{ij}c^2 \ll R$, we obtain 
\begin{equation}
\delta \rho(u,v)\sim -c^2\sim -e^{-2\ell_G/\xi}.
\end{equation}

\subsection{Network Energy}
\label{app:energy}
The dissipated energy for a tube $(i,j)$, given a sink--source couple $(s, t)$ is
$$\mathcal{E}_{ij}(s,t)=R_{ij}q_{ij}(s,t)^2=g_{ij}(p_i-p_j)^2.$$

Summing over all (undirected) edges,
\begin{align*}
\mathcal{E}(s,t) &=\sum_{(i,j)\in E}\mathcal{E}_{ij}(s,t)
=\sum_{(i,j)\in E}g_{ij}(p_i-p_j)^2 \\
&=\vec{p}^TL\vec{p}
=\vec{p}^T\vec{b}_{st}
=\vec{b}_{st}^TL^\dagger\vec{b}_{st}
=R_{\mathrm{eff}}(s,t),
\end{align*}
where \(R_{\mathrm{eff}}(s,t)\) is the effective resistance between \(s\) and \(t\). Summing over all unordered source--sink pairs gives the total network dissipation,
\begin{equation}
\mathcal{E}_{\mathrm{tot}} =\sum_{s<t}\mathcal{E}(s,t) =\sum_{s<t}R_{\mathrm{eff}}(s,t) =n\,\operatorname{Tr}(L^\dagger) =Kf(G),
\end{equation}
where $Kf(G)$ is the Kirchhoff index. This result was proven in \cite{Tetali1991} and \cite{Ghosh2008}, we give it here for completeness. Here we use the decomposition of the effective resistance $R_{\rm{eff}}(s, t) = (L^\dagger)_{ss}+(L^\dagger)_{tt}-2(L^\dagger)_{st}$, which gives $\mathcal{E}_{\rm{tot}} =\sum_{s<t}\left[(L^\dagger)_{ss}+(L^\dagger)_{tt}-2(L^\dagger)_{st}\right]$. The diagonal terms ($ss$ and $tt$) give $\sum_{s<t}\left[(L^\dagger)_{ss}+(L^\dagger)_{tt}\right]=(n-1)\sum_{s}(L^\dagger)_{ss} = (n-1)\rm{Tr}(L^\dagger)$.
For the off-diagonal terms, we use the constraint $L^\dagger \cdot \vec{1}^T=0$, i.e. the rows and columns sum to zero: $\forall i, \sum_{j}(L^\dagger)_{ij}=0$. Thus, $\sum_{j\neq i}(L^\dagger)_{ij} = -(L^\dagger)_{jj}$. Summing over $i$, $\sum_{ij}(L^\dagger)_{ij}=0 \Leftrightarrow\sum_{i\neq j}(L^\dagger)_{ij} = -\sum_{i}(L^\dagger)_{ii} = -\rm{Tr}(L^\dagger)$. Finally, since $\sum_{i\neq j}(L^\dagger)_{ij}=2\sum_{i<j}(L^\dagger)_{ij}$, we find the off diagonal terms to be $2\sum_{s<t}(L^\dagger)_{st}=-\rm{Tr}(L^{\dagger})$. 

Summing diagonal and non-diagonal terms lead to $$\mathcal{E}_{\rm{tot}} = (n-1)\rm{Tr}(L^\dagger) + \rm{Tr}(L^\dagger) = n \rm{Tr}(L^\dagger)$$

\subsection{Change in total dissipation due to edge pruning}
To derive the change in total dissipation, $\delta \mathcal{E}_{\rm{tot}}=\mathcal{E}'_{\rm{tot}}-\mathcal{E}_{\rm{tot}}$, upon removing edge $(i,j)$ we start by noting that  $\mathcal{E}_{\rm{tot}} = n\text{Tr}(L^{\dagger})$ and $\mathcal{E}'_{\rm{tot}} = n\text{Tr}(L'^{\dagger})$, such that by linearity of the trace $\delta \mathcal{E}_{\rm{tot}} = n\text{Tr}(\delta L^{\dagger})$. Since $\delta L^\dagger = \gamma_{ij} L^\dagger \vec{b}_{ij}\otimes \vec{b}_{ij}^T L^\dagger$, we find
\begin{equation}
    \delta \mathcal{E}_{\rm{tot}} = n\frac{1+\rho_{ij}}{\rho_{ij}R_{ij}}S_{ij},
\end{equation}
recalling $\gamma_{ij}= (R_{ij}-R_{\rm{eff}}(i,j))^{-1}= \frac{1+\rho_{ij}}{R_{ij} \rho_{ij}}$ and defining $S_{ij}=\text{Tr}\left(L^\dagger \vec{b}_{ij}\otimes \vec{b}_{ij}^T L^\dagger\right) = \Vert L^\dagger \vec{b}_{ij} \Vert^2$. We refine the term $S_{ij}$ further by defining $p_k(i,j) = \vec{e}_k^T\cdot\vec{p}(i,j)$, with $\vec{p}(i,j) = L^\dagger \vec{b}_{ij}$ the vector field using $i$ as source, $j$ as a sink. Thus, $S_{ij} = \Vert L^\dagger \vec{b}_{ij}\Vert^2 = \Vert \vec{p}(i, j)\Vert^2=  \sum_k p_k^2(i,j)$. Now, using the identity $\sum_{s<t} (x_s-x_t)^2 = n \sum_k x_k^2 - \left(\sum_k x_k\right)^2$ with $x_s=p_s(i,j)$ and $x_t=p_t(i,j)$, and noting that, since $L^\dagger 1 = 0$, $\sum_k p_k (i,j)=0$, thus, $\sum_{s<t} \left[p_s(i,j)-p_t(i,j)\right]^2 = n \sum_k p_k^2(i,j) = n S_{ij}$.
Finally, we express the pressure drop $p_s(i,j)-p_t(i,j)=\tilde{q}_{st}(i,j)$ with the flow through a resistance one edge connecting node $i$ and $j$ to find that 
$S_{ij} = \frac{1}{n}\sum_{s<t}\tilde{q}^2_{st}(i,j)$, and thus a convex function. 

\subsection{Network sparsification bounds}
\label{app:sparsification}
The Spielman-Srivastava algorithm produces a \textbf{sparse network} $H$, obtained from a network $G$, such that the Laplacian $L_H$ is bounded by an error parameter $\varepsilon$ as
$$\forall \vec{x}\in\mathbb{R}^n, (1-\varepsilon)\vec{x}^T \, L_G\, \vec{x} \leq \vec{x}^T \, L_H\, \vec{x} \leq (1+\varepsilon)\vec{x}^T \, L_G\, \vec{x}$$
In our case, when removing an edge $(i,j)$, we obtain the bounding:
\begin{equation}
    \left(1-\frac{1}{1+\rho_{ij}}\right)\vec{x}^TL_G\vec{x} \leq \vec{x}^TL_H \vec{x}\leq \vec{x}^T L_G \vec{x}
\end{equation}

The upper-bound is obtained since $L_H = L_G - g_{ij}\vec{b}_{ij}\otimes\vec{b}_{ij}$. Applying $\vec{x}$ on both sides: 
\begin{align*}
\vec{x}^TL_G\vec{x} &= \vec{x}^T L_H\vec{x}+ g_{ij} \vec{x}^T \cdot \vec{b}_{ij}\otimes \vec{b}_{ij}^T\cdot\vec{x} \\
&=  \vec{x}^T L_H\vec{x} + g_{ij}(\vec{b}_{ij}^T\cdot \vec{x})^2 \\
&= \vec{x}^TL_H\vec{x} + g_{ij}(x_i-x_j)^2
\end{align*}
Then since $g_{ij}(x_i-x_j)^2\geq0$, the RHS is always positive and we get $\vec{x}^TL_H\vec{x}\leq \vec{x}^TL_G\vec{x}$.

The lower bound is obtained as follow: since $\vec{1}^T\cdot\vec{b}_{ij}=0$, then $(L_G^\dagger)^{1/2}(L_G)^{1/2}\vec{b}_{ij}=\vec{b}_{ij}$, thus, applying Cauchy-Schwartz on $\vec{b}^T_{ij}\cdot \vec{x}=((L_G^\dagger)^{1/2}\vec{b}_{ij})^T(L_G^{1/2}\vec{x})\equiv\vec{u}^T\cdot \vec{v}$ leads to
$$(\vec{u}^T\cdot \vec{v})^2\leq (\vec{u}^T\cdot \vec{u})(\vec{v}^T\cdot \vec{v}) \Leftrightarrow (\vec{b}_{ij}\cdot \vec{x})^2 \leq (\vec{b}_{ij}^TL_G^\dagger \vec{b}_{ij})(\vec{x}^T L_G\vec{x})$$
Since $\vec{b}_{ij}\cdot\vec{x}=x_i-x_j$ and by definition, $R_{\rm{eff}}(i,j)=\vec{b}_{ij}^TL_G^\dagger\vec{b}_{ij}$, we have
$$(x_i-x_j)^2 \leq  R_{\rm{eff}}(i,j)\vec{x}^TL_G\vec{x}$$
thus, $g_{ij}(x_i-x_j)^2 \leq g_{ij}R_{\rm{eff}}(i,j) \vec{x}^TL_G\vec{x}$. Finally, since $\vec{x}^TL_G\vec{x}-\vec{x}^TL_H\vec{x}=g_{ij}(x_i-x_j)^2$, we obtain
\begin{align*}
\vec{x}^TL_G\vec{x}-\vec{x}^TL_H\vec{x} &\leq g_{ij}R_{\rm{eff}}(i,j)\vec{x}^TL_G\vec{x} \\
\Leftrightarrow \left[1-g_{ij}R_{\rm{eff}}(i,j)\right]\vec{x}^TL_G\vec{x} &\leq \vec{x}^TL_H\vec{x}\end{align*}

Recalling $g_{ij}R_{\rm{eff}}(i,j)=\frac{1}{1+\rho_{ij}}$, we obtain the bounding.

\subsection{Change in Fiedler value due to edge pruning}
\label{app:fiedler}
Inspired by \cite{Cai2023a, Luo2025} we derive the change in Fiedler value due to pruning edge $(i,j)$. We denote the second eigenvector/value of the changed graph Laplacian $L'=L+\delta L$ with $\delta L=-R_{ij}^{-1}\vec b_{ij}\vec b_{ij}^T$ as $\vec v_2'=\vec v_2+\delta\vec v_2$ with $\lambda_2'=\lambda_2+\delta\lambda_2$. Substituting into $L'\vec v_2'=\lambda_2'\vec v_2'$ and using $L\vec v_2=\lambda_2\vec v_2$ we find,
\begin{equation}
L\,\delta\vec v_2+\delta L(\vec v_2+\delta\vec v_2)=\lambda_2\,\delta\vec v_2+\delta\lambda_2(\vec v_2+\delta\vec v_2). 
\label{eqn_fiedler1}
\end{equation}
Projecting onto $\vec v_2$ gives, exactly,
\begin{equation}
\delta\lambda_2=\underbrace{\vec v_2^T\delta L\,\vec v_2}_{A}+\underbrace{\vec v_2^T\delta L\,\delta\vec v_2}_{B}. 
\end{equation}
Introducing $d_k\equiv\vec b_{ij}^T\vec v_k=v_k(i)-v_k(j)$, we get $A=-g_{ij}d_2^2$. From the pseudo-inverse identity $R_{\rm eff}(i,j)=\vec b_{ij}^TL^\dagger\vec b_{ij}=\sum_{k\ge2}d_k^2/\lambda_k$ and $g_{ij}R_{\rm eff}(i,j)=(1+\rho_{ij})^{-1}$ it follows,
\begin{equation}
A=-\frac{\lambda_2}{1+\rho_{ij}}+g_{ij}\lambda_2\!\sum_{k\ge3}\frac{d_k^2}{\lambda_k}=-\frac{\lambda_2}{1+\rho_{ij}}+O\left(g_{ij}\frac{\lambda_2}{\lambda_3}\right), 
\label{eqn_fiedlerA}
\end{equation}
using $\lambda_k\ge\lambda_3$. To estimate $B$ we project Eq.~\eqref{eqn_fiedler1} onto $\vec v_k$, $k\neq2$, the $k=1$ term vanishes identically since $\vec{v}_1=1^T/\sqrt{N}$, $d_1=\vec b_{ij}^T\vec v_1=0$, leaving $k\ge3$. To leading order, $\vec v_k^T\delta\vec v_2\simeq-g_{ij}d_kd_2/(\lambda_2-\lambda_k)$, so $\delta\vec v_2=-g_{ij}d_2\sum_{k\ge3}d_k\vec v_k/(\lambda_2-\lambda_k)+O(\text{h.o.t.})$, and thus
\begin{equation}
B=g_{ij}^2d_2^2\sum_{k\ge3}\frac{d_k^2}{\lambda_2-\lambda_k}.
\label{eqn_fiedlerB}
\end{equation}
Since $\lambda_k-\lambda_2\ge\lambda_3-\lambda_2$ and $\sum_kd_k^2=\|\vec b_{ij}\|^2=2$ (Parseval's identity),  we bound $|B|\le 2g_{ij}^2d_2^2/(\lambda_3-\lambda_2)$; and since $g_{ij}d_2^2\leq \lambda_2 g_{ij}R_{\rm eff}(i,j)=\lambda_2/(1+\rho_{ij})\le\lambda_2$ (as $\rho_{ij}\ge0$, i.e.~$R_{ij}\ge R_{\rm eff}(i,j)$),
\begin{equation}
|B|\leq\frac{2g_{ij}\lambda_2}{\lambda_3-\lambda_2}=O\left(g_{ij}\frac{\lambda_2}{\lambda_3}\right).
\end{equation}
Combining the result for A Eq.~\eqref{eqn_fiedlerA} and B Eq.~\eqref{eqn_fiedlerB} we find,
\begin{equation}
\delta\lambda_2=A+B=-\frac{\lambda_2}{1+\rho_{ij}}+O\!\left(g_{ij}\frac{\lambda_2}{\lambda_3}\right),
\end{equation}
that the Fiedler value decays shallower for larger $\rho_{ij}$

\section{Numerical implementation}

\subsection{Relative resistance from Kron reduction}
The quantity $R_{\mathrm{net}}(u,v)$ can be computed efficiently through Kron reduction \cite{Dorfler2013}. Kron's reduction eliminates all interior nodes by partitioning the set of nodes $V$ as an interior set $I = \lbrace i \in V | i \neq u,v\rbrace$, while preserving the electrical response at the boundary $B=\lbrace u,v\rbrace$. As a result, the entire remainder of the network is replaced by a single equivalent resistance $R_{\mathrm{net}}(u,v)$ connecting nodes $u$ and $v$. Formally, Kron's reduction reads:
$$L_{\rm{red}} = L_{BB} - L_{BI}L^{-1}_{II}L_{IB}$$
where $L_{\rm{red}}$ is the reduced Laplacian, $L_{II}$ (resp. $L_{BB}$) is the sub-matrix with only interior (resp. boundary) nodes, and $L_{IB}, L_{BI}$ are the cross-elements. As we specifically choose $(u, v)\in E$, the reduced Laplacian off-diagonal elements directly correspond to the inverse network resistance. Note that here, we only need to inverse $L_{II}$, which is non-singular. While this is a costly operation, we do not need to inverse it at every iterations of the pruning schemes, since we can use Sherman-Morrison formula to calculate the new inverse from the previous one.

\emph{Implementation.} \texttt{Function\_RelResKronRed\_v2.m} builds the sparse Laplacian once, then loops over the $m$ edges. For each edge it subtracts $g_{uv}$ from the four entries of $L$ and solves $L_{II}^{-1}L_{IB}$ with a sparse solve.

\subsection{Network generation and pruning rules}

\paragraph{Regular graphs}. We generate regular graphs based on regular 2-dimensional lattices. The topology of the graph (hexagonal, triangular, square) and the number of layers are the main parameters. The regular graphs can be periodic.

\paragraph{Poisson-Delaunay graphs}. Poisson-Delaunay are generated from a spatial Poisson distribution of centers, which generates nodes around the centers as a gaussian distribution. Edges of the graph are then generated using a Delaunay triangulation. The main parameters of these graphs are the box size (spatial extent of initial Poisson distribution), the number of clusters, and the cluster spread (spatial spread of each cluster).

\paragraph{Erdos--Renyi graphs.} The number of nodes $N_0$ and probability $p$ are determined arbitrarily, to generate graphs of comparable size to the regular ones. The probability $p$ is chosen such that the graphs are connected, typically $p>0.2$.

For all graphs, we decide if the edges have identical resistance, or if it is noisy, thus the resistance is given by a random normal value.

\paragraph{Pruning rules}. At each iteration exactly one edge is removed, selected by one of four rules:
\begin{itemize}
\item \textbf{Maximum relative ressistance:} $e^\star=\arg\max_{e\in E} \rho_e$.
\item \textbf{Random:} $e^\star$ drawn uniformly among all edges.
\item \textbf{Boltzmann($T$):} this is inspired from Boltzmann-weighted softmax sampling. We associate to each edge a probability $p(e)\propto \exp[(\rho_e-\max_{e'}\rho_{e'})/T]$, which recovers greedy for $T\to0$ and random for $T\to\infty$. To choose an edge, we build the cumulative distribution function (CDF) from the normalized probabilities, and we draw a random number u in the [0,1] interval and select the first edge in the CDF that reaches u.
\end{itemize}
A candidate whose removal disconnects the graph is rejected. For maximum relative resistance, the next best edge is tried ($\rho_e\to-\infty$), while for random and Boltzmann the draw is repeated. The procedure stops when $\max_e\rho_e\le 1$, i.e.\ when every remaining edge carries at least half of the current for a unit injection across its own endpoints. This criterion is the same for all rules.

\subsection{Pruning loop and data storage}

For every realization, one graph is generated and shared by all rules. The routine \texttt{pruneGraphSchemes(G,mode,T)} then iterates:
\begin{enumerate}
\item compute $\rho$ (via Kron reduction), $L^{\dagger}=\mathrm{pinv}(L)$, $E_{\rm tot}$ and $R_{\rm eff}$ for every edge;
\item store the current state in \texttt{History(t)} (fields \texttt{R}, \texttt{rho}, \texttt{G}, \texttt{Etot});
\item stop if $\max_e\rho_e\le1$;
\item select an edge with the chosen rule, reject it if it disconnects the graph, otherwise remove it.
\end{enumerate}
\texttt{History(t)} is the state \emph{before} the $t$-th removal. Its length is therefore the number of removed edges plus one, the last entry is the final graph, and step $t$ has $m-t+1$ edges.

The results are saved in two files per ensemble. The compact file \texttt{ensemble\_*.mat} contains \texttt{rho\_all\{c,r\}\{t\}}, \texttt{R\_all\{c,r\}\{t\}}, \texttt{Etot\_all\{c,r\}(t)}, \texttt{avgDegree\_all\{c,r\}(t)}, \texttt{nEdges\_all\{c,r\}(t)}, \texttt{mode\_all}, \texttt{T\_all} and \texttt{nNodes}, with $c$ the rule (maximum relative resistance, random, Boltzmann), and $r$ the realisation. The file \texttt{pruning\_graphs\_*.mat} contains \texttt{graphs.pos} and, for each rule, \texttt{graphs.<rule>\{r\}\{t\}} with the edge list, $\rho$, $R$, $E_{\rm tot}$ and the numbers of nodes and edges. The rule name is the mode, or \texttt{boltzmann\_T<100\,T>} for Boltzmann (for example \texttt{boltzmann\_T25} for $T=0.25$).

For the lattices all resistances are equal unless specified and thus use an initial gaussian distribution of edge resistances, so maximum relative resistance runs are deterministic and identical across realizations. Their variability comes from the random and Boltzmann draws.

\section{Supplementary Figures}
\begin{figure}[h]
    \centering
    \includegraphics[width=0.5\linewidth]{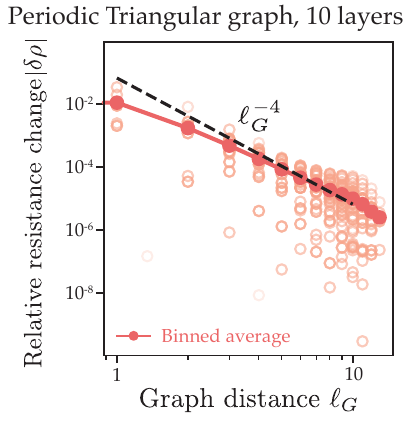}
    \caption{\textbf{Change of relative resistance upon edge removal scales similarly for periodic boundary networks as non-periodic networks}. Periodic boundaries triangular network with 10 layers are used, uniform resistance $R_{ij}=1$. The dashed line is drawn as a guide to the eye.}
    \label{suppfig:relreschange_PBC10layers}
\end{figure}

\begin{figure}[h]
    \centering
    \includegraphics[width=1.\linewidth]{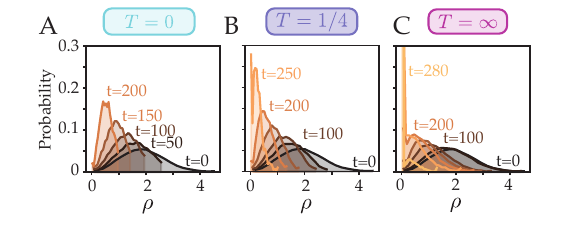}
    \caption{\textbf{Relative resistance distribution for three temperatures of pruning schemes}. We use a triangular network with 5 layers, noisy initial resistances $R_{ij}=1 \pm \varepsilon$, $\varepsilon=0.3$, non-periodic boundary conditions. We indicate with a color code different iterations steps. 50 simulations are ensemble-averaged. We truncate the y-axis for better visibility.}
    \label{suppfig:relativeresistance_distribution}
\end{figure}

\begin{figure}[h]
    \centering
    \includegraphics[width=0.5\linewidth]{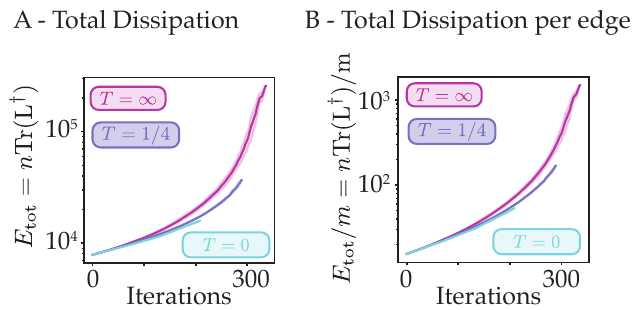}
    \caption{\textbf{Maximum relative resistance pruning leads to networks with low total dissipation}. (A) The total dissipation and (B) total dissipation rescaled by the number of edges show that the maximum relative resistance schemes reache lower values than higher temperature schemes.}
    \label{suppfig:dissipationenergy}
\end{figure}

\begin{figure}[h]
    \centering
    \includegraphics[width=0.5\linewidth]{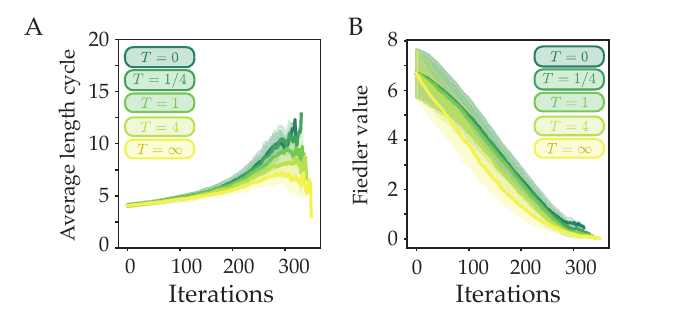}
    \caption{\textbf{Average length cycle and Fiedler value for Erdős–Rényi networks with iterations}. Full lines are the average, shaded area the standard deviation. We used parameters N=50, p=0.3. 50 simulations are ensemble-averaged.}
    \label{suppfig:randomnetworks_morphology}
\end{figure}


\end{document}